\documentclass[twocolumn,prl,final]{revtex4}
\usepackage[T1]{fontenc}
\usepackage[latin9]{inputenc}
\usepackage{graphicx}
\usepackage{hyperref}
\usepackage{amssymb}

\begin{document}

\title{Saddle-node bifurcation cascade in optically injected lasers}

\author{Jesús San Martín }

\affiliation{Dep. Matemática Aplicada, EUITI - Universidad Politécnica de Madrid,
Ronda de Valencia 3, 28012 Madrid, Spain}

\author{Daniel Rodríguez-Pérez}

\affiliation{Dep. Física Matemática y de Fluidos, UNED, Senda del Rey 9, Madrid,
Spain}

\begin{abstract}
Self-similar structures converging to accumulation points are observed
in optically injected lasers. These structures are associated to saddle-node
bifurcations. We found that these phenomena can be properly explained
in the framework of saddle-node bifurcation cascade. Apparently not
connected phenomena in these lasers are shown to be related within
this framework.
\end{abstract}


\maketitle
Lasers with optical reinjection have played an important role in recent
years, from three different points of view, namely theoretical, numerical
and experimental. Recently, the presence of self-similar arrangements
of periodic orbits has been reported in this class of lasers \citep{Kranskopf2002,Bonatto2007,Wieczorek2001,Wieczorek2005,Yeung1998}.
The phenomena described in the literature are associated to saddle-node
(S-N) bifurcations and to periodic orbits accumulation points. Furthermore,
smaller regions in parameter space have been reported showing the
same behavior (see \S 3.2 in \citep{Kranskopf2002}). Also the presence
of accumulation points of accumulation points has been described \citep{Bonatto2007}.

A new mathematical framework is needed to organize most or all of
these results, or rephrasing \citep{Bonatto2007}, to answer the question
of what is the precise structure of the laser chaotic phases. In this
letter we will show this framework: the saddle-node bifurcation cascade
\citep{SanMartin2007a,SanMartin2007b}. Formerly, we will show how
the diverse phenomena described by researchers fit in this framework
in a qualitative way. Later, after introducing the mathematical model
of laser, we will show it in a quantitative way. This new framework
would make the analysis of the problem easier, differentiating between
basic processes and secondary effects. In fact, this model explains
the following phenomena:

\paragraph{(i) Self-similar structures.}

The self-similar structures above mentioned (associated to S-N bifurcations,
with a fractal structure and converging to accumulation points that,
on the other hand, converge to accumulation points), have the same
structure as that found in the S-N bifurcation cascade, which was
fully characterized in \citep{SanMartin2007a,SanMartin2007b}.

\paragraph{(ii) Accumulation of infinite cascades.}

A S-N bifurcation cascade is a sequence of S-N bifurcations in which
the number of fixed points showing this kind of bifurcation is duplicated.
If the whole canonical window is considered, then infinitely many
S-N bifurcation cascades exist, formed by sequences of points with
periodicities $q$, $q\cdot2$, $q\cdot2^{2}$, \ldots{}, $q\cdot2^{n}$,
\ldots{}, for any $n\in\mathbb{N}$ and $q$ odd, called basic period
(see definition in \S3.2 of \citep{SanMartin2007a}). These sequences
converge, as $n$ grows, to the same point (in a way resembling that
of the accumulation horizons reported in \citep{Bonatto2007}). This
point is known as the canonical window Myrberg-Feigenbaum point. Each
of the $q\cdot2^{n}$-periodic S-N orbits gives birth to a $q\cdot2^{n}$-periodic
window inside which all the process is repeated again as it was in
the canonical window, hence giving a fractal structure. Hence each
of these $q\cdot2^{n}$-windows has its own Myrberg-Feigenbaum accumulation
point, and all these accumulation points converge toward the canonical
window Myrberg-Feigenbaum accumulation point. In the injected laser
literature this is reported by Bonatto and Gallas \citep{Bonatto2007}
as {}``boundaries formed by the accumulation of infinite cascades
of self similar islands of periodic solutions of ever-increasing period''
and {}``\ldots{}these chaotic phases contain both single accumulations
as well as accumulations of accumulations''. The kind of phenomena
described by Bonatto and Gallas matches perfectly what is called {}``attractor
of attractors'' in \citep{SanMartin2007a}, as we will see later
in more detail.

\paragraph{(iii) Unnested period-doubling structures.}

In \citep{Wieczorek2001}, Wieczorek, Kranskopf and Lenstra describe
unnested period-doubling structures in injected lasers. They say:
{}``the next bifurcation is the saddle-node of the periodic orbits
$SL^{2}$ in which two period-$2$ orbits, one attracting and one
saddle are created''. It is likely that they are describing the two
first elements of a S-N bifurcation cascade, that is, an unnested
island of doubled period seems to be one element in a saddle node
bifurcation cascade. In other words, the S-N bifurcation gives birth
to a window that appears as an unnested island.

\paragraph{(v) Intermittency.}

Wieczorek, Kranskopf and Lenstra also report chaotic behavior close
to the S-N bifurcation \citep{Wieczorek2001}. They are observing
one single element of the intermittency cascade associated to the
S-N bifurcation cascade \citep{SanMartin2007a}. In summary, before
the S-N bifurcation, there is intermittency, and after it the system
enters a periodic window where the initial periodic orbit undergoes
a period-doubling cascade leading to chaos. We will show this intermittency
below. Within the theoretical framework just described, the S-N bifurcation
is the main element and the secondary is the intermittency which is,
in fact, naturally associated to it.

It is enough to find a S-N bifurcation cascade in reinjected lasers
to apply the theoretical framework just described to these systems.
In this letter, we find S-N bifurcation cascades in the same reinjected
laser model used by \citep{Kranskopf2002,Bonatto2007,Wieczorek2005,Wieczorek2001},
providing the mathematical basis to explain the phenomena described
above. We have chosen this laser model because periodic orbits associated
to S-N bifurcations have been found in it which agree with experimental
results. Furthermore, the model is quite general and representative
of a broad number of class-B lasers (solid state and $\mbox{C}\mbox{O}_{2}$
lasers), as pointed out in \citep{Kranskopf2002}. In this way, we
expect to ease the experimental research based in our theoretical
developments. The equations of the model can be written in a dimensionless
form as\[
\dot{E}=K+(\frac{1}{2}(1+i\alpha)n-i\omega)E\]
\[
\dot{n}=-2\Gamma n-(1+2Bn)(\vert E\vert^{2}-1)\]
(see \citep{Kranskopf2002} for details); the parameter values have
been taken as $\alpha=1.987$, $\omega=1.5$, $\Gamma=0.035$ and
$B=0.015$. 

\begin{figure}[htbp]
\begin{centering}
\includegraphics[width=0.95\columnwidth]{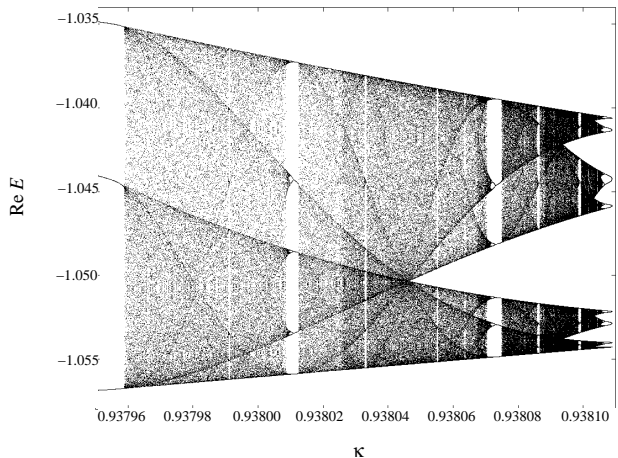}
\par\end{centering}

\caption{\label{fig:5window}Fragment of the period-$5$ window. Several periodic
windows are seen within it. The origin of each of these periodic windows
(its right bound) coincides with the birth of saddle-node orbits of
the corresponding saddle-node bifurcation cascade.}

\end{figure}

Taking $\kappa$ as the control parameter, S-N bifurcation cascades
are found (see figure \ref{fig:5window}). In tables \ref{tab:sn3},
\ref{tab:sn5} and \ref{tab:sn5x3}, the values of $\kappa$ where
S-N orbits of several periods appear are shown, as well as the Feigenbaum
ratios, converging to the Feigenbaum constant $\delta$. They show
the $3\cdot2^{n}$ and $5\cdot2^{n}$ cascades as well as the S-N
bifurcation cascade within the period-$5$ window. This proves, in
optically injected lasers:

\paragraph*{(a) The existence of cascades with different basic periods ($3$
and $5$).}

This explains the existence of unnested islands.

\paragraph*{(b) The same kind of behavior in smaller regions (S-N bifurcation
cascade in the period-$5$ window).}

This explains the existence of self-similar structures.

\paragraph*{(c) The underlying attractor of attractors originated by the presence
of cascades inside each of the windows of the cascade.}

This explains the accumulation of accumulation points.

Then, the phenomena reported by researchers are explained within the
S-N bifurcation cascade framework.

Let us point out that all of the S-N bifurcation cascades have the
same scaling as the Feigenbaum cascade \citep{Feigenbaum1978,Feigenbaum1979}
allowing this property for a technique to have them characterized
by researchers (see tables \ref{tab:sn3}, \ref{tab:sn5} and \ref{tab:sn5x3}). 

Another point has to be highlighted, the $q$-periodic orbit, for
$q$ odd, is located in the $1$-chaotic band, the $q\cdot2$, lies
in the $2$-chaotic band, and so on. This allows us to explain the
structures described by Bonatto and Gallas. For instance, orbits $3$,
$5$, $7$ (the last one not numbered but shown) in Fig. 1b in \citep{Bonatto2007},
belong to the $1$-chaotic orbit. The elements located in the 2-chaotic
band allow to explain the low periodic islands with periods $10$,
$14$, $18$, $22$ shown in Fig. 2b in \citep{Bonatto2007}. They
are low periodic islands of periods $5\cdot2$, $7\cdot2$, $9\cdot2$
and $11\cdot2$ which are respectively associated with the second
elements of the S-N bifurcation cascade $q\cdot2^{n}$, having basic
periods $q=5,\;7,\;9\;\mbox{y}\;11$. The periods of the islands just
mentioned follow a Sharkovsky ordering \citep{Sharkowskii1964}, which
claims that the low periodic island of period $6$ (shown in figure
1b in \citep{Bonatto2007}) is in reality a $3\cdot2$-periodic island,
that is, it belongs to the $2$-chaotic band and represents the second
element of the S-N bifurcation cascade with basic period $q=3$. Closeups
of these periodic islands can be seen in the biparametric plot in
figure \ref{fig:unnested-island} (similar to figure 1b in \citep{Bonatto2007}).
If the biparametric structure is visited along a straight line across
the unnested periodic islands, then a plot similar to the one in figure
\ref{fig:5window} is obtained from which the S-N bifurcation cascades
can be extracted.

In this same framework, it is possible to explain the two kinds of
convergence reported by researchers:

\paragraph*{(d) Simple accumulation.}

If for $q\cdot2^{n}$, $n$ is kept fixed and $q$ is increased, the
corresponding sequence represented by the low periodic islands in
the $2^{n}$-chaotic band converges to the Misiurewicz or band-merging
point where two consecutive bands meet each other \citep{SanMartin2007b}.
This is seen in Fig. 2a in \citep{Bonatto2007} where $n=1$ and $q=5,\:7,\:9,\:11,\:\dots$
and the low periodic islands converge to an {}``accumulation horizon''.

\paragraph*{(e) Accumulation of accumulations.}

If, for a fixed $q$, $n$ is increased, the corresponding sequence
converges to an attractor of attractors (see \citep{SanMartin2007a}).

In fact, all the convergence points are attractors of attractors when
observed in enough detail, because within any window there exists
an infinity of attractors of attractors. However the software or experimental
setups employed cannot resolve the simple accumulation and therefore
the accumulation of accumulations cannot always be observed.

\begin{figure}[htbp]
\begin{centering}
\includegraphics[width=0.95\columnwidth]{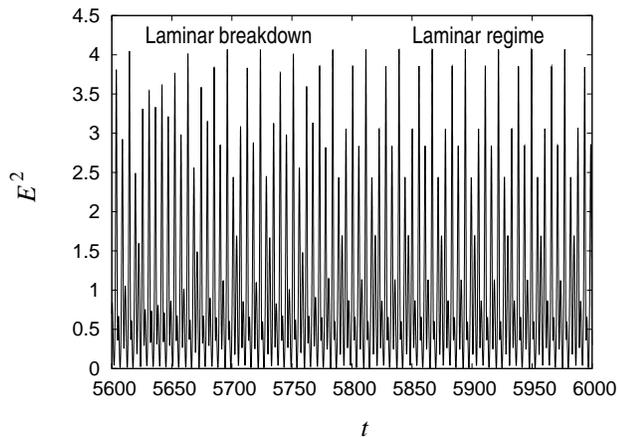}
\par\end{centering}

\caption{\label{fig:Intermittency}Intermittency behavior close to the onset
of the $5$-periodic window, for a value of $\kappa=0.938818$.}

\end{figure}

\begin{figure}[htbp]
\begin{centering}
\includegraphics[width=0.95\columnwidth]{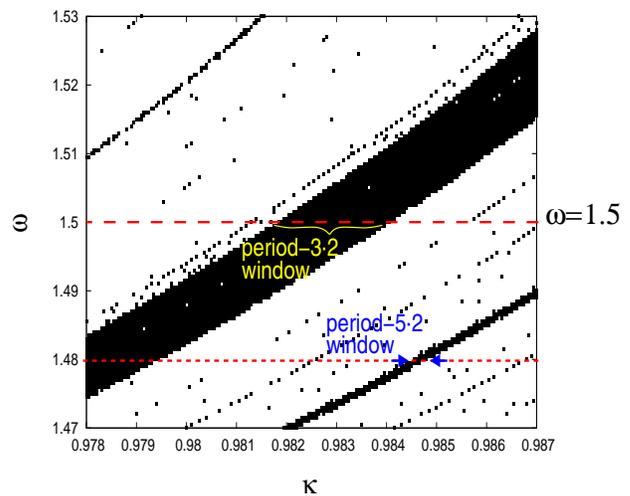}
\par\end{centering}

\caption{\label{fig:unnested-island}Low periodic islands of different periods
($3\cdot2$ and $5\cdot2$) born at S-N bifurcations and located in
the $2$-chaotic band. These S-N bifurcations belongs to different
S-N bifurcation cascades (basic periods $3$ and $5$ respectively).
The points in this plot have been computed by period-counting.}

\end{figure}

Finally, there must be an intermittency before any S-N bifurcation.
In particular, \citep{Wieczorek2001} report an intermitency behavior.
But, in general, there must be intermittency surounding every low
$q\cdot2^{n}$-periodic periodic island described above. These would
be responsible of the appearance of intermittency cascades de intermitencias
behaving in a similar way as S-N bifurcation cascade do. We show one
of these intermittencies in figure \ref{fig:Intermittency}.

According to the exposed in the previous paragraph, to show up the
existence of a S-N bifurcation cascade in a dynamical system the chaotic
band containing every $q\cdot2^{n}$-periodic orbit (for a fixed $q$
basic period) of the cascade has to be perfectly identified. Otherwise,
that is, if the succesive windows that appear when the control parameter
is varied are registered in order, then the S-N bifurcation cascade
will not be readily obtained, but a mixture of unrelated periodic
windows, that do not follow the Feigenbaum scaling any more. The S-N
bifurcation cascades are subsequences of this sequence that have to
be chosen carefully according to the (symbolic) rules given in \citep{SanMartin2007a}.

\begin{table}[htbp]
\begin{tabular}{|c|c|c|}
\hline 
period & $\kappa_{n}$ & $(\kappa_{n-1}-\kappa_{n-2})/(\kappa_{n}-\kappa_{n-1})$\\
\hline
\hline 
$3\times2$ & $0.984222$ & \\
\hline 
$3\times2^{2}$ & $0.994836$ & \\
\hline 
$3\times2^{3}$ & $0.997134$ & $4.619$\\
\hline 
$3\times2^{4}$ & $0.997621$ & $4.719$\\
\hline 
$3\times2^{5}$ & $0.997726$ & $4.638$\\
\hline
\end{tabular}

\caption{\label{tab:sn3}Values of $\kappa$ corresponding to S-N bifurcation
cascade of period $3$ in the canonical window.}

\end{table}

\begin{table}[htbp]
\begin{tabular}{|c|c|c|}
\hline 
period & $\kappa_{n}$ & $(\kappa_{n-1}-\kappa_{n-2})/(\kappa_{n}-\kappa_{n-1})$\\
\hline
\hline 
$5$ & $0.938815$ & \\
\hline 
$5\times2$ & $0.989430$ & \\
\hline 
$5\times2^{2}$ & $0.995914$ & $7.806$\\
\hline 
$5\times2^{3}$ & $0.997363$ & $4.475$\\
\hline 
$5\times2^{4}$ & $0.997671$ & $4.705$\\
\hline
\end{tabular}

\caption{\label{tab:sn5}Values of $\kappa$ corresponding to S-N bifurcation
cascade of period $5$ in the canonical window.}

\end{table}

\begin{table}[htbp]
\begin{tabular}{|c|c|c|}
\hline 
period & $\kappa_{n}$ & $(\kappa_{n-1}-\kappa_{n-2})/(\kappa_{n}-\kappa_{n-1})$\\
\hline
\hline 
$(5\times3)$ & $0.937958$ & \\
\hline 
$(5\times3)\times2$ & $0.9380750$ & \\
\hline 
$(5\times3)\times2^{2}$ & $0.9380993$ & $4.815$\\
\hline 
$(5\times3)\times2^{3}$ & $0.93810450$ & $4.673$\\
\hline
\end{tabular}

\caption{\label{tab:sn5x3}Values of $\kappa$ corresponding to S-N bifurcation
cascade of period $3$ inside the period-$5$ window.}

\end{table}

In summary, all kind of phenomena (self-similar structures, accumulation
points, unnested period-doubling structures, intermittency and so
on) reported by researchers in optically injected lasers can be explained
in the theoretical framework of saddle-node bifurcation cascade. Furthermore,
new unreported phenomena (intermittecy cascade) have been predicted.

\end{document}